\documentclass[%
 reprint,
 amsmath,amssymb,
 aps,
prab,superscriptaddress,showkeys
]{revtex4-2}

\usepackage{graphicx}
\usepackage{dcolumn}
\usepackage{bm}

\begin{document}


\title{\textbf{Simulation model of crystal-based beam extraction using BDSim toolkit enhanced with Geant4 G4ChannelingFastSimModel} 
}%

\author{Alexei Sytov}\email{sytov@fe.infn.it}
\affiliation{INFN Ferrara Division, Via Saragat 1, Ferrara, 44122, Italy}

\author{Gero Kube}
\affiliation{Deutsches Elektronen-Synchrotron DESY, Notkestr. 85, Hamburg, 22607, Germany}

\author{Laura Bandiera}
\affiliation{INFN Ferrara Division, Via Saragat 1, Ferrara, 44122, Italy}

\author{Vincenzo Guidi}
\affiliation{INFN Ferrara Division, Via Saragat 1, Ferrara, 44122, Italy}
\affiliation{Department of Physics and Earth Science, University of Ferrara, Via Saragat 1, Ferrara, 44122, Italy}

\author{Andrea Mazzolari}
\affiliation{INFN Ferrara Division, Via Saragat 1, Ferrara, 44122, Italy}
\affiliation{Department of Physics and Earth Science, University of Ferrara, Via Saragat 1, Ferrara, 44122, Italy}

\author{Gianfranco Paternò}
\affiliation{INFN Ferrara Division, Via Saragat 1, Ferrara, 44122, Italy}

\author{Sergey Strokov}
\affiliation{Deutsches Elektronen-Synchrotron DESY, Notkestr. 85, Hamburg, 22607, Germany}

\date{\today}

\begin{abstract}
We present a simulation model for crystal-based beam extraction of lepton beams from an accelerator by using a bent crystal. This model utilizes the BDSim toolkit for particle dynamics in an accelerator in conjunction with the Geant4 G4ChannelingFastSimModel for beam deflection in a bent crystal and G4BaierKatkov for radiation losses. In order to link these codes, we designed a new BDSim accelerator component - a bent crystal compatible with G4ChannelingFastSimModel.

As a demonstration, we constructed a complete simulation model of the DESY II Booster Synchrotron within BDSim, incorporating all relevant accelerator components and apertures. The model accounts for betatron and synchrotron oscillations, as well as radiation losses in an oriented crystal, using the Baier-Katkov radiation integral. Simulation results demonstrate the extraction of a primary monoenergetic 6 GeV electron beam, characterized by a charge of 23 pC, a horizontal beam emittance of 3.6 $\mu$m$\cdot$rad, a vertical beam emittance of 0.32 $\mu$m$\cdot$rad, and an energy spread of 0.008. Under the simulated conditions, (1.42 $\pm$ 0.02 \%) of the initial
circulating beam reaches the septum entrance. This corresponds to an
extraction efficiency of (28.1 $\pm$ 0.4 \%) when evaluated relative to the
particles intercepted by the bent crystal.

The approach provides a powerful tool to significantly accelerate R\&D for applications related to crystal-based extraction and collimation in lepton accelerators and colliders.
\end{abstract}

\maketitle


\section{Introduction}
\label{introduction}

\begin{figure}
	\centering 
	\includegraphics[width=0.46\textwidth]
    {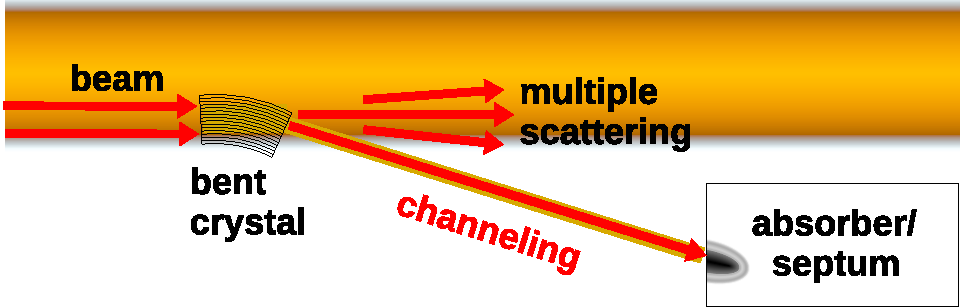}	
	\caption{Schematic representation of a crystal-based collimation/extraction for the particle beam deflected into an absorber/septum magnet, respectively.}
	\label{Fig0}
\end{figure}

Channeling in bent crystals \citep{Tsyganov} offers an efficient method for extracting particle beams from accelerators. Charged particles entering a bent crystal at small angles are guided by its atomic planes, enabling high-quality beam extraction for fixed-target experiments, high-energy lepton collider studies, and crystal-based collimation as illustrated in Fig. \ref{Fig0}. These techniques have been tested at proton and ion synchrotrons, including U70, RHIC, SPS, Tevatron, and LHC \citep{U70,RHIC,UA92,Tevatron,LHC,LHCion}, with dedicated SPS studies demonstrating the importance of optimizing the crystal and absorber settings in a multi-turn collimation system \citep{SPS2}. However, such techniques have not yet been applied to electron machines.

Our previous studies demonstrated the feasibility of a proof-of-principle experiment on the extraction of 6 GeV electrons from the DESY II Booster Synchrotron \citep{DESYII} using crystal-based techniques. These methods enable monoenergetic beam extraction without disrupting primary experiments, making them ideal for various applications like detector R\&D \citep{Diener}, and beyond-Standard Model physics. This approach is adaptable to existing and future electron, positron and muon synchrotrons, linacs and colliders, offering broad research potential.

\begin{figure*}
	\centering 
	\includegraphics[width=0.95\textwidth]
    {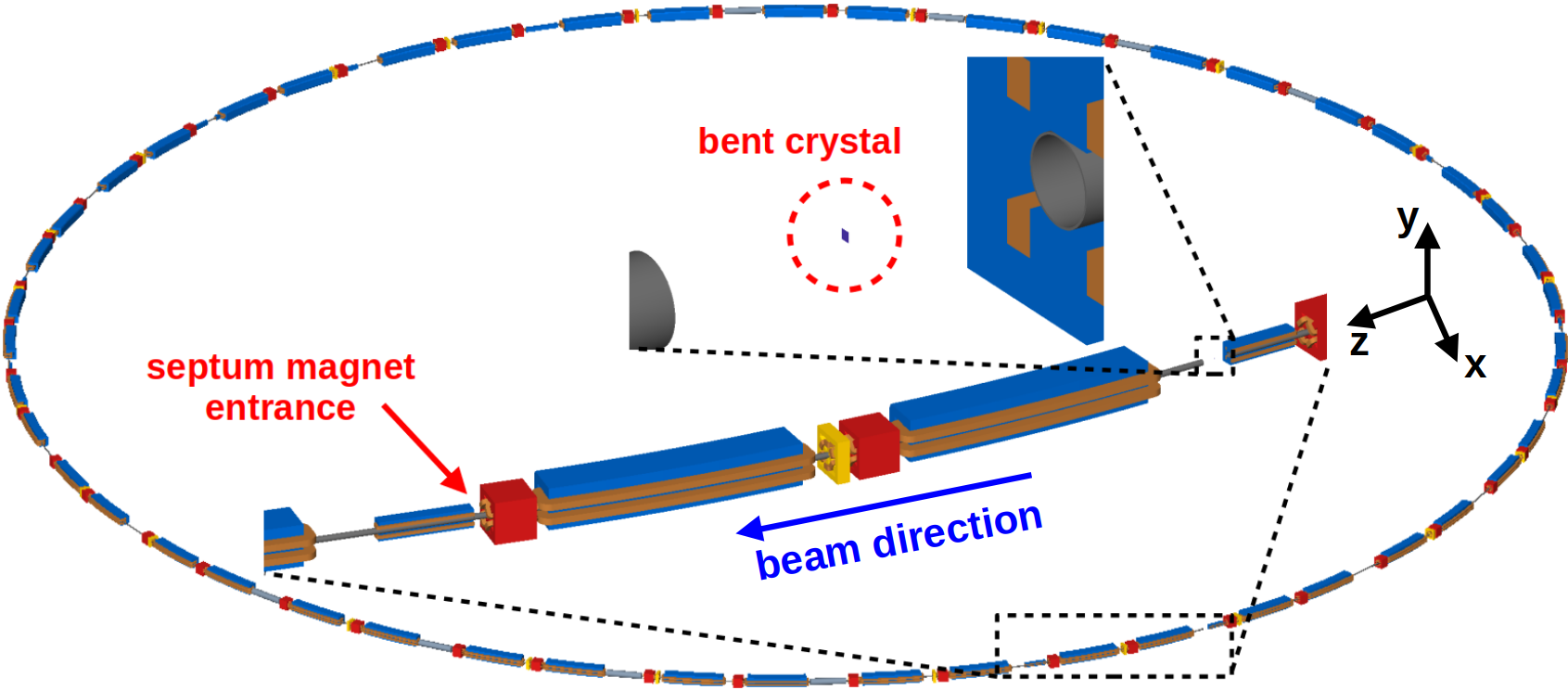}
	\caption{BDSim visualization of the DESY II Booster Synchrotron model with a crystal-based extraction setup. The beam direction and the coordinate system are schematically indicated.}
	\label{Fig1}
\end{figure*}

However, the complexity of the extraction process necessitates detailed simulation tools capable of modeling particle dynamics through various accelerator components, crystal interactions, energy losses and synchrotron oscillations. In particular, at high energies, the radiation losses in a crystal may become critical due to synchrotron oscillations, since radiation losses under the channeling conditions significantly exceed bremsstrahlung by intensity due to coherent effects of radiation - channeling radiation \citep{Kumakhov,Baryshevsky,Baier} and coherent bremsstrahlung \citep{Ferretti,Uberall,Ter-Mikaelian,Baier}.

\textbf{BDSim} \citep{BDSim1,BDSim2,BDSim3}, a versatile simulation tool based on the \textbf{Geant4} \citep{Geant41,Geant42} toolkit, simulates particle transport and interactions in accelerators, offering precise thick-lens tracking and integrating electromagnetic, hadronic, and user-defined Geant4 physics models. This code has already been used for the simulations of crystal-based collimation exploiting \textbf{G4Channeling} model \citep{G4Channeling, BDSimCollimation}. However, this channeling model is adapted for high energy protons/ions and includes neither precise simulation of suppression of incoherent scattering in a crystal lattice \citep{CSS} nor coherent effects of radiation mentioned above, both of which affect the deflection of leptons at $\sim$ GeV energy. Other simulation models for crystal-based extraction/collimation such as models within Sixtrack and Fluka \citep{Sixtrack,SixtrackFluka} also possess the same limitations.

These missing coherent processes were implemented into Geant4 within another model - \textbf{G4ChannelingFastSimModel} and \textbf{G4BaierKatkov} describing channeling and radiation processes, respectively \citep{Sytov,CRYSTALRAD,G4Ch,Riccardo}. Being available since the Geant4 v11.2.0 official release, \textbf{G4ChannelingFastSimModel} and \textbf{G4BaierKatkov} have already been used successfully in simulations of real case experiments on particle deflection \citep{Sytov,PRL2025}, radiation and crystal-based positron source \citep{Nicola} demonstrating good agreement with experimental data.  For example, recent measurements with a 530 MeV positron beam resolved a fine structure in the angular distribution of particles emerging from a bent crystal, providing a stringent experimental benchmark for models of particle dynamics in bent crystals \citep{PRL2025}. Together with BDSim, these tools enable complete simulations of crystal-based extraction and collimation of a lepton beam, being essential to design real accelerator applications.

In this study, we integrate the BDSim framework with Geant4 containing \textbf{G4ChannelingFastSimModel} and \textbf{G4BaierKatkov}, by creation of a bent crystal as a BDSim user defined accelerator component compatible with these models. We implement the DESY II Booster Synchrotron based on our previous studies \citep{DESYII} and carry out a detailed simulation of the crystal-based extraction of 6 GeV electron beam. Additionally, the key parameters of the extracted beam, including charge, transverse emittance, and energy spread as well as the extraction efficiency, are calculated to demonstrate the advantages of this technique for future applications. Furthermore, we discuss the applicability of the model for crystal-based extraction/collimation at future colliders and demonstrate the importance of correct simulations of coherent effects of radiation in a crystal.

\section{Implementation of a bent crystal into BDSim}
\label{BDSim}

\subsection{Physics description}

\subsubsection{BDSim}

The \textbf{BDSim} toolkit \citep{BDSim1,BDSim2,BDSim3} utilises Geant4 as a library to simulate particle transport and interaction with accelerator materials. It includes a wide collection of accelerator components, such as magnets, cavities, kickers etc. the parameters of which can be fully set up from input files including the geometry of accelerator apertures and electromagnetic fields. A graphical representation of a full accelerator ring built in BDSim through Geant4 Graphical User Interface (GUI) is shown in Fig. \ref{Fig1} (its description is given in the next section).

\subsubsection{G4ChannelingFastSimModel}

\begin{figure}
	\centering
	\includegraphics[width=0.46\textwidth]
    {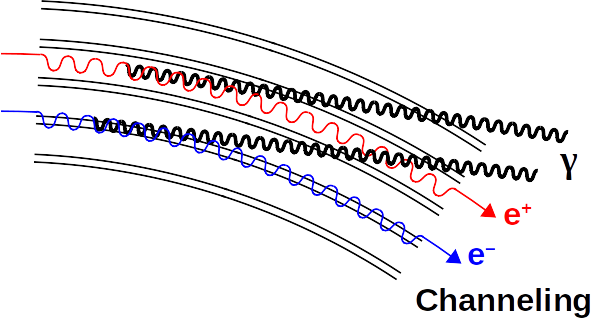}
	\caption{Schematic illustration of channeling of positrons and electrons and radiation production in a bent crystal formed by non-equidistant crystal planes (e.g. (111) planes of a silicon crystal).}
	\label{Fig15}
\end{figure}

\textbf{G4ChannelingFastSimModel} \citep{Sytov,CRYSTALRAD,G4Ch,Riccardo} is a Geant4 model that accurately simulates the interaction of charged particles with oriented crystals, including coherent effects such as channeling. Channeling in a crystal \citep{Lindhard,Tsyganov}, schematically represented in Fig. \ref{Fig15}, is an effect of charged particle penetration through a monocrystal moving almost parallel to crystallographic planes (planar channeling) or axes (axial channeling) and oscillating in their transverse electric field. This field is capable of confining a charged particle around the transverse equilibrium position similarly to a focusing field in an accelerator but at Angstrom level precision. In this paper only planar channeling will be simulated. However, there is no limitation to exploit axial effects as well, which will be briefly discussed at the end of the paper.

In Fig. \ref{Fig15} one can also observe a clear distinction between electron and positron channeling. Due to their opposite charge, positrons oscillate between crystal planes, whereas electrons cross the planes at each oscillation. This increases multiple scattering for electrons and consequently reduces their channeling fraction compared to positively charged particles. Fig. \ref{Fig15} illustrates the case of (111) non‑equidistant silicon planes (the primary configuration considered in this paper) which differs from (110) equidistant planes.

The angular acceptance of the channeling effect is described by the so-called critical channeling angle (Lindhard angle) as:

\begin{equation}
\theta_{L}=\sqrt{\frac{2 U_0}{pv}},
\label{Eq01}
\end{equation}
where $U_0$ is a potential well depth, p and v being particle momentum and velocity.

Channeling effect is possible not only in a straight but also in a bent crystal \citep{Tsyganov}, if its radius exceeds a critical value, which can be calculated as:

\begin{equation}
R_{cr}=pv/E_{max},
\label{Eq02}
\end{equation}
where $E_{max}$ is a maximal transverse electric field.

An accurate simulation of the channeling effect requires:

\begin{itemize}

 \item detailed simulation of a particle trajectory in the transverse electric field taking into account the crystal geometry in the case of a bent crystal;

 \item simulation of incoherent scattering on separate atoms taking into account its partial suppression due to the contribution of coherent scattering on crystal planes/axes \citep{CSS};

 \item taking into account the transverse distribution of nuclear and electron density;

 \item simulation of ionization losses in a crystal lattice.

\end{itemize}

All of these effects were implemented into the \textbf{G4ChannelingFastSimModel}. More information about the model and its validation vs experimental data can be found in dedicated papers, such as: \citep{Sytov,Riccardo,PRL2025,Nicola,PositronSource}.

\subsubsection{G4BaierKatkov}

For light particles i.e. electrons and positrons radiation losses become crucial. Moreover, at multi-GeV energies a quantum recoil effect, when a particle loses a significant amount of its energy by emission of a single photon, is also of high importance. These effects are taken into account in so-called Baier-Katkov radiation integral \citep{Baier}, which is taken over a classical trajectory such as calculated using \textbf{G4ChannelingFastSimModel}.

The Baier-Katkov integral was implemented into Geant4 within \textbf{G4BaierKatkov} class used by
\textbf{G4ChannelingFastSimModel} to simulate a secondary photon emission and radiation losses as well.

\subsection{Implementation mechanism}

To integrate an oriented bent crystal into BDSim, which allows for user-defined accelerator components, one must define both the component class and its constructor. This is achieved by creating two classes: the component itself, which is part of the Geant4 DetectorConstruction, and the constructor that reads component parameters from a macro and creates the component class.

We developed \textbf{CrystalDeflector} as the component class and \textbf{CrystalDeflectorConstructor} as its constructor. The constructor handles parameter setup for the crystal, including dimensions, material properties, offsets, orientation, and specific parameters for crystal channeling and radiation modeling.

\textbf{CrystalDeflector} creates the crystal volume according to the parameters above and activates \textbf{G4ChannelingFastSimModel} for simulating the channeling effects and \textbf{G4BaierKatkov} for radiation losses within the defined crystal G4Region, i.e. the list of Geant4 Logical Volumes in which the model is active \citep{Geant41,Geant42}.

To incorporate these models into the simulation, Geant4's Fast Simulation Interface must be used. This interface is enabled in the BDSim user-defined Physics List for specific particles (e- and e+ in the case of lepton machines). After compiling the relevant classes, the accelerator simulation setup can be created and added to the input macro files following the standard BDSim procedure described in BDSim user's guide \citep{BDSim3}. The user-defined component (\textbf{CrystalDeflector}) must be listed and appropriately positioned within the accelerator lattice in the input files. After these steps, the simulation is ready to be executed.

The corresponding macro command for the crystal considered in this paper is described in Appendix \ref{appendix}. For more information please refer the user's guide \citep{GITHUB}.

\section{Simulation setup}
\label{setup}

\begin{table}
\caption{\label{Table1}DESY II parameters for the simulation setup}
\begin{ruledtabular}
\begin{tabular}{l c} 
 Parameter &Value\\
 \hline
Ring circumference $S_0$& 292.8 m\\
Nominal extraction energy $E_0$& 6 GeV\\
Number of $e^{-}$ in the beam $N_0$&  $\sim 10^{10}$\\
Horizontal emittance $\varepsilon_x$ (at 6 GeV)& 339 nm$\cdot$rad\\
Vertical emittance $\varepsilon_y$ (at 6 GeV)& 35 nm$\cdot$rad\\
Energy spread $\sigma_{\delta E/E_0}$& $0.977\times10^{-3}$\\
RF voltage per cavity $V_s$& 0.9875 MV\\
Total number of RF cavities & 8\\
RF cavity frequency $h$& 499.654 MHz\\
RF cavity phase & $\pi$\\
Septum magnet boundary & 0.98 mm (4)\\
Septum magnet boundary thickness & 3 mm
\end{tabular}
\end{ruledtabular}
\end{table}

The simulation setup for the DESY II Booster Synchrotron was created by converting DESY II MAD-X files into the BDSim format. Beam and lattice parameters from previous studies \citep{DESYII} were retained (Table \ref{Table1}). Unlike the earlier setup, this setup simulates the entire beam distribution rather than only the halo, and tracks it along the experimental ring, thereby naturally providing the ab-initio distribution of the beam at the crystal entrance. Moreover, unlike our previous studies, this BDSim configuration includes all accelerator components, with magnets modeled as thick lenses and beam apertures and RF cavities accurately positioned along the ring. This approach naturally accounts for betatron and synchrotron oscillations. The cavities were phased for storage ring operation, while synchrotron radiation losses were neglected, since in reality they are routinely compensated by RF acceleration and minimally impact the extraction process. The total cavity voltage was adjusted based on the actual parameters. The visualization of the full DESY II ring as well as of the extraction zone is shown in Fig. \ref{Fig1}.

\begin{figure}
	\centering
	\;\;\includegraphics[width=0.46\textwidth]
    {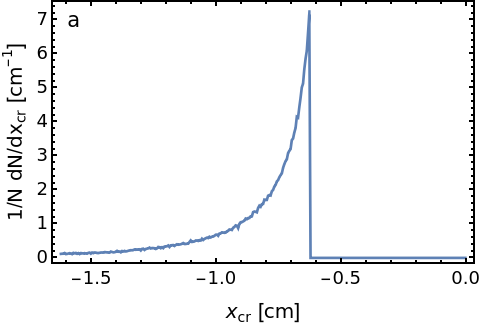}
    \includegraphics[width=0.46\textwidth]
    {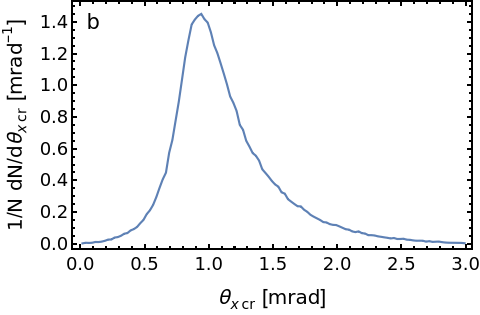}
	\caption{Horizontal coordinate (impact parameter) (a) and horizontal angular (b) distributions of the particles hitting the crystal.}
	\label{Fig2}
\end{figure}

\begin{table}
\caption{\label{Table2}Bent crystal parameters for the simulation setup}
\begin{ruledtabular}
\begin{tabular}{l c}
 Parameter &Value\\
 \hline
Material & silicon\\
Crystallographic plane& (111)\\
Thickness & 175 $\mu$m\\
Bending angle & 1.75 mrad\\
Boundary transverse position & -0.63 cm (-3$\sigma$) \\
Angular alignment & 0.97 mrad
\end{tabular}
\end{ruledtabular}
\end{table}

The parameters of the bent crystal optimized in earlier studies \citep{DESYII} are presented in Table \ref{Table2} (see also \ref{appendix}). Coherent effects of radiation in the crystal were taken into account. 

Particles that intercept the crystal may be deflected by channeling (and by other coherent/incoherent processes modeled by \textbf{G4ChannelingFastSimModel}) and can then reach the
septum entrance within the tracked turns, 100 in our case. A BDSim beam dump was placed at the location of the septum magnet entrance to record particles arriving at this point and terminate their tracking upon arrival.

We have utilized Geant4 v11.2.1 and the FTFP\_BERT physics list. The setup is, however, fully compatible with newer Geant4 releases, and we recommend using the latest Geant4 version.

\begin{figure*}
	\centering
	\begin{tabular}{cc}
	\includegraphics[width=0.47\textwidth]
    {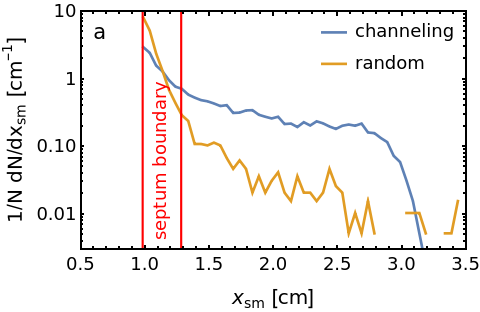} &
    \includegraphics[width=0.44\textwidth]
    {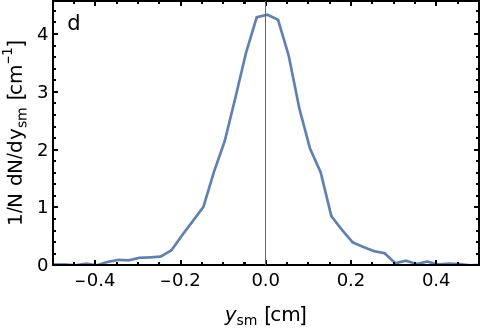}\\

    \includegraphics[width=0.45\textwidth]
    {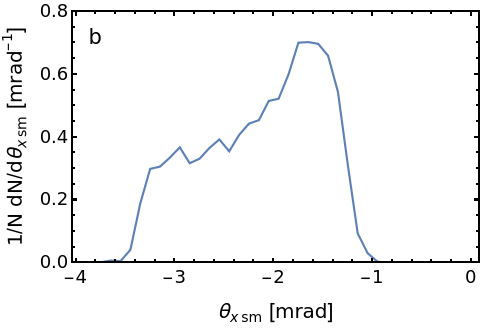} &
    \includegraphics[width=0.45\textwidth]
    {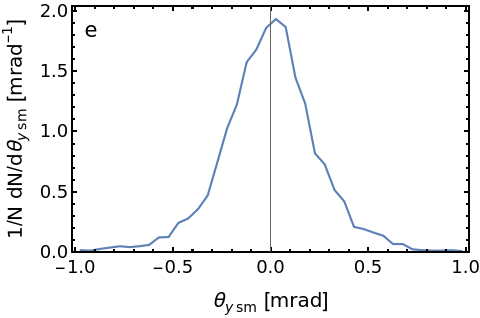}\\

    \includegraphics[width=0.5\textwidth]
    {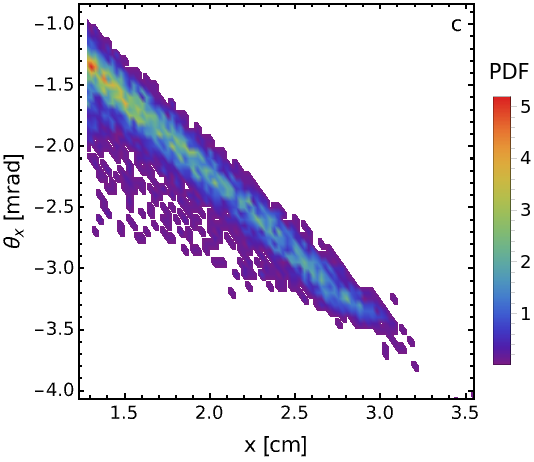}	&
    \includegraphics[width=0.51\textwidth]
    {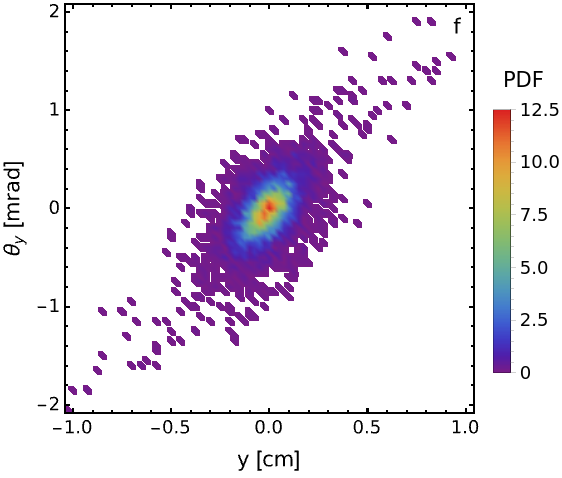}
    \end{tabular}
	\caption{The distributions of the horizontal coordinate (a), the angle (b) and phase space (c) as well as the same distributions in vertical direction (d-f, respectively) of particles entering the septum magnet (excluding its boundary for b-f). (a) also shows the case of random crystal alignment.}
	\label{Fig3}
\end{figure*}

\begin{figure}
	\centering
	\includegraphics[width=0.46\textwidth]
    {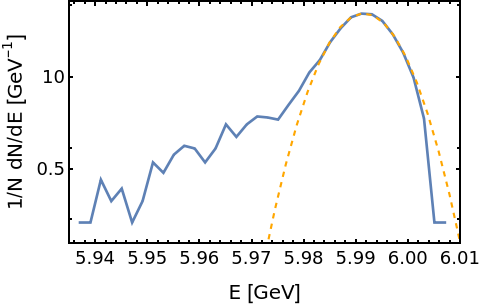}

	\caption{The energy distribution of particles entering the septum magnet (excluding its boundary). The dashed curve shows the Gaussian fit of the distribution peak.}
	\label{Fig35}
\end{figure}

\section{Simulation results}
\label{results}

Figure \ref{Fig2} illustrates the impact parameter distribution of particles at the crystal and their horizontal angular distribution, counting also multiple arrivals of the same particles. Most particle hits occur within the crystal’s boundary (placed at -3$\sigma \approx -6.3$ mm of the beam), corresponding to the peak of the angular distribution at the angle $\theta_{x \; peak}=-\alpha (-3\sigma)/\beta \approx 0.97$ mrad (see Table \ref{Table2}), where $\alpha$ and $\beta$ are the Twiss parameters. A slight shift of the peak position (Fig. \ref{Fig2}b) can be explained by synchrotron oscillations and energy losses in the crystal before the subsequent crystal passage. This alignment determined the crystal orientation in our setup, though further adjustments are possible. Particles are typically captured under channeling conditions if they approach within the critical channeling angle, which for 6 GeV electrons and (111) silicon planes equals to 88 $\mu$rad (for straight crystal planes), according to Eq. (\ref{Eq01}). A more precise estimate taking in account the bending crystal radius to be higher than the critical radius (see Eq. (\ref{Eq02})) with a factor of 9, in our case, slightly reduces the critical channeling angle down to 77 $\mu$rad. However, one should also consider the non-channeling particles that can be captured under channeling conditions due to multiple scattering inside the crystal \citep{rechanneling,rechanneling2}, which is taken into account during simulations.

Figure \ref{Fig3} presents the simulated distributions of the coordinates and angles of the extracted beam and its phase spaces at the septum magnet entrance both in horizontal and vertical direction. The simulations consumed $\sim$1.5 kcore-hours of computational time to generate the statistical sample from an initial beam of 390k electrons.

The results are consistent with those from earlier studies \citep{DESYII}, demonstrating similar distributions. However, it is important to stress that the present simulations employ a more complete and accurate model featuring a full accelerator lattice, refined synchrotron oscillation dynamics, a complete aperture map, the full beam distribution (rather than only the halo), and up-to-date versions of the channeling and Baier-Katkov models. Fig. \ref{Fig3}a also provides a comparison with the randomly oriented crystal case, when the amount of extracted particles is nearly 18 times less than for channeling.

Figure \ref{Fig35} shows the energy distribution of the extracted beam, which remains highly monochromatic. A small tail is present due to radiation losses in the crystal. However, unlike prior results, this tail is minimized due to the realistic aperture simulated in BDSim, which removes particles that escape the accelerator's RF bucket.

From these distributions, the key extracted beam parameters including horizontal and vertical transverse emittance, charge, r.m.s. energy spread, and extraction efficiency were calculated and are summarized in Table \ref{Table3}.

\begin{table}
\caption{\label{Table3}Simulation results: parameters of the extracted beam}
\begin{ruledtabular}
\begin{tabular}{l c}
 Parameter &Value\\
 \hline
Horizontal emittance $\epsilon_{x\,extr}$ & 3.6 $\mu$m$\cdot$rad\\
Vertical emittance $\epsilon_{y\,extr}$ & 0.32 $\mu$m$\cdot$rad\\
Beam energy (peak) & 5.99 GeV \\
R.m.s. energy spread (Gaussian fit) & 0.005 \\
R.m.s. energy spread (std dev) & 0.008 \\
Beam charge & 23 pC\\
Total beam fraction & 1.42 $\pm$ 0.02 \%\\
Extraction efficiency & 28.1 $\pm$ 0.4 \%
\end{tabular}
\end{ruledtabular}
\end{table}

The transverse emittance of particles entering the septum is calculated according to
\begin{equation}
\epsilon_{x\,extr}=\sqrt{\langle x^2 \rangle \langle x'^2 \rangle-\langle x \cdot x'\rangle^2},
\label{Eq03}
\end{equation}
where $x$ and $x'$ denote the particle positions and angles, respectively. The vertical emittance $\epsilon_{y\,extr}$ is calculated analogously.

The peak energy and its r.m.s. energy spread were calculated using a Gaussian fit as demonstrated in Fig. \ref{Fig35}. In order to take into account a non-Gaussian tail the r.m.s. energy spread was also recalculated as a standard deviation value.

The total beam fraction is defined as the fraction of extracted particles relative to the total beam population (1.42 $\pm$ 0.02 \%), whereas the extraction efficiency is the fraction relative only to those particles intercepted by the bent crystal (28.1 $\pm$ 0.4 \%). The latter value is higher than the
approximately 16 \% obtained in the earlier study \citep{DESYII}. This difference is likely related to the more complete description of the beam dynamics, the use of the full beam distribution, and the detailed mapping of the accelerator apertures in the present BDSim model, having a particular impact on the beam distribution at the crystal entrance. Since the two simulation setups differ in their beam sampling and aperture treatment, this comparison should be regarded as indicative rather than as a direct performance comparison.

These results demonstrate the benefits of crystal-based extraction of a primary beam with low emittance, high charge, and minimal energy spread. It is important to note that in our setup, approximately $\sim$ 95 \% of the beam never interacted with the crystal and was, therefore, unaffected. Furthermore, all extracted beam parameters can be adjusted by further optimizing the crystal's transverse position, alignment, and geometry, as well as by exploiting effects beyond channeling, which we briefly discuss in the following section.

\section{Future applications}
\label{applications}

\begin{figure*}
	\centering
	\begin{tabular}{cc}
	\includegraphics[width=0.45\textwidth]
    {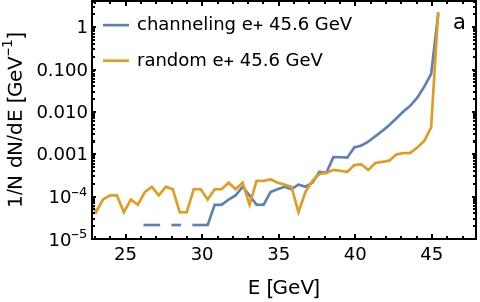} &
    \includegraphics[width=0.45\textwidth]
    {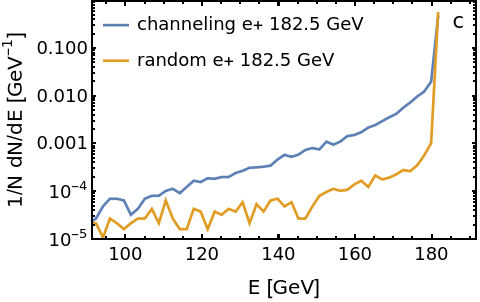}\\

    \includegraphics[width=0.45\textwidth]
    {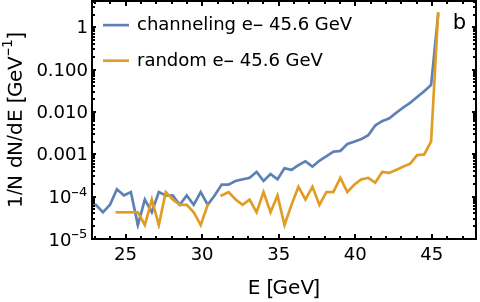} &
    \includegraphics[width=0.45\textwidth]
    {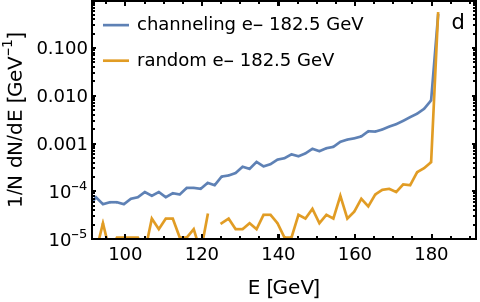}
    \end{tabular}
	\caption{Energy distribution of the beam after its interaction with the silicon bent crystal for positrons (a,c), electrons (b,d) at the energy 45.6 GeV (a,b) and 182.5 (c,d). The simulation parameters are listed in Table \ref{Table4}.}
	\label{Fig4}
\end{figure*}

The simulation model developed in this work is intended for accelerator systems in which an oriented bent crystal is used as a compact beam-steering element, while the crystal interaction must be treated consistently with particle transport through the surrounding lattice. \textit{Future applications} of this model potentially include the development of crystal-based extraction and collimation techniques at a wide range of existing and future lepton accelerators, colliders, synchrotron light sources, and linac-driven facilities such as XFELs. Furthermore, because the model includes coherent radiation emission in oriented crystals, it can also provide a basis for assessing applications in which radiation production becomes central.

These potential applications can be broadly categorized into the following groups:

\begin{itemize}

 \item \textbf{Test facilities} for nuclear and particle physics detector R\&D, electronics irradiation tests for space applications, etc. \citep{Applications1} at existing synchrotrons and linacs, with the energies ranging from a few hundred of MeV, e.g. DA$\Phi$NE collider \citep{Dafna}, up to few GeV at synchrotron light sources and their boosters, similar to the DESY case considered in this paper \citep{ApplicationsDESY, DESYII}.

 \item \textbf{Crystal-based collimation at future colliders} such as FCC-ee \citep{BDSimCollimation} and CEPC and potentially muon colliders \cite{Muon}, requiring circulation of extremely high intensity and very high energy beams.

 \item \textbf{Crystal-based extraction at future colliders} for various fixed-target experiments beyond the collider physics. In particular, slow extraction of 20–46 GeV positrons from the FCC-ee booster has been considered as a route to an NA64-like experiment exploring new regions of light-dark-matter parameter space \citep{beyondFCC}. In addition, the use of electron and positron beams for studies of spin dynamics in oriented crystals under the channeling conditions, including measurements of spin precession and tests of strong-field modifications of the anomalous magnetic moment \citep{BaryshevskySpin,TikhomirovSpin,TikhomirovPolar}, has recently been proposed for FCC-ee \citep{beyondFCC}. This application is directly related to the present work, although its simulation would require the inclusion of spin transport and spin-dependent processes in addition to the channeling and radiation models employed here.

 \item \textbf{Coherent source of gamma radiation and secondary particles in crystals} 
 
 Coherent effects of radiation can be exploited to produce intense quasi-monochromatic gamma radiation. The effect of coherent bremsstrahlung produced by electrons in a crystal has already been successfully exploited for intense linearly polarized gamma-ray source to probe nuclear and hadron physics at Mainz Mikrotron MAMI \citep{CBMainz} and JLAB \citep{CBJLAB}. Various crystal-based radiation sources were explored at MAMI \citep{ApplicationsMainz,PRL2015,EPJC2021} and SLAC FACET Facility \citep{SLAC2017,SLAC2019} with both electron and positron beams. A further extension concerns radiation-source applications based on periodically bent oriented crystals, so-called crystalline undulators \citep{Baryshevsky,CU2,CU3,CU4}. In this configuration, channeling positrons follow periodically deformed crystal planes and can generate intense hard X-ray or gamma-ray radiation which can also be exploited at future facilities including FCC-ee \citep{beyondFCC}.  Furthermore, as mentioned above, oriented crystals can be exploited for a positron source for future colliders such as FCC-ee, also requiring detailed simulation of beam transport \citep{PositronSource,PositronSource2}.
 
\end{itemize}

An example of the first application was simulated in this paper, i.e. crystal-based extraction of electrons — a typical case for synchrotrons and synchrotron light sources. This approach can be readily extended to linear accelerators, the key difference being that the linac case does not require multi-turn simulations, thus simplifying the development process.

Besides planar channeling in a bent crystal, other effects can compete in the efficiency of deflecting charged particles. These include axial channeling \citep{Lindhard} and stochastic deflection \citep{Axial1,Axial2}, which exploits an axial field exceeding that of planar channels by approximately an order of magnitude, and multiple volume reflection from different crystallographic planes in a single bent crystal \citep{MVROC,MVROC2,PRA2012}, first observed experimentally with high-energy proton beams \citep{MVROCCERN}. Additionally, crystal materials beyond silicon -- such as germanium \citep{rechanneling2} -- can be considered to further optimize performance. All these physical scenarios can be set up within the \textbf{G4ChannelingFastSimModel}.

The second and third applications (and potentially also a radiation source - the fourth application) involve higher-energy beams at future colliders, ranging from 45-182.5 GeV at FCC-ee and CEPC, up to the sub‑TeV and TeV scale at the ILC and CLIC. A circular collider can potentially benefit from crystal-based collimation system as was already considered in \citep{BDSimCollimation}, while crystal-based extraction is a promising technique for both circular and linear machines.

While channeling efficiency would improve at higher energies due to reduced multiple scattering, a key difference from the GeV case studied in this paper is the importance of radiation losses. Indeed, as can be observed in Fig. \ref{Fig35} and as previously studied \citep{DESYII} the radiation losses in our 6 GeV case do not significantly affect the process of the crystal-based extraction. However, the energy increase also causes the emission mechanism to shift from dipole-like to synchrotron-like radiation, leading to a significant rise in hard photon production and correspondingly fast radiation losses \citep{PRL2018,PRA2012}.
To demonstrate this effect at FCC-ee energies, we performed single-pass simulations of energy loss in a bent crystal using the \textbf{G4ChannelingFastSimModel}. The simulations considered an ideal channeling crystal orientation and a low-divergence monoenergetic beam for both positrons and electrons at the minimum and maximum FCC-ee energies -- 45.6 and 182.5 GeV. The parameters, based on earlier studies \citep{BDSimCollimation}, are listed in Table \ref{Table4}. In our setup, the crystal geometry and crystal planes differed between electrons and positrons but were kept fixed as the energy increased, while the angular divergence was maintained the same and well below the critical channeling angle across the entire energy range.

\begin{table}
\caption{\label{Table4}Simulation parameters used to evaluate radiation-induced energy losses in a silicon bent crystal at FCC-ee beam energies.}
\begin{ruledtabular}
\begin{tabular}{l c}
 Parameter &Value\\
 \hline
Material & silicon\\
Crystallographic plane for e-& (111)\\
Crystallographic plane for e+& (110)\\
Thickness for e- & 100 $\mu$m\\
Thickness for e+ & 200 $\mu$m\\
Bending angle & 100 $\mu$rad\\
Angular divergence & 1 $\mu$rad
\end{tabular}
\end{ruledtabular}
\end{table}

The simulation results are shown in Fig. \ref{Fig4}. A pronounced low-energy tail of particles that have lost a significant fraction of their energy is clearly visible. This effect is already considerable at the lowest FCC‑ee energy and becomes even more significant at higher energies. Electron radiation losses exceed those of positrons due to stronger scattering with atoms, since electrons oscillate around crystal planes during channeling, whereas positrons channel between them, as illustrated in Fig. \ref{Fig15}.

\begin{figure}
	\centering
	\includegraphics[width=0.45\textwidth]
    {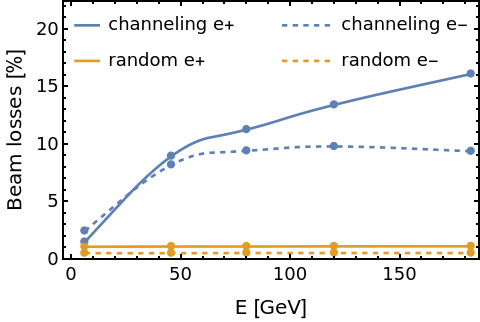}

	\caption{
    Fraction of particles emerging from the silicon bent crystal with an energy loss larger than 1 \% of the incident energy at 6 GeV and at FCC-ee beam energies. Points show the simulated values, while lines represent interpolation. The simulation parameters are listed in Table \ref{Table4}.}
	\label{Fig5}
\end{figure}

To provide a common indicator of radiation-induced beam degradation, we evaluated the fraction of particles emerging from the crystal with an energy loss exceeding 1 \% of the incident energy. This threshold is used here as a representative benchmark rather than as the exact momentum acceptance of a specific machine, which depends on the lattice configuration and on the crystal location. Simulations were performed at
all FCC-ee beam energies (45.6, 80, 120, and 182.5 GeV), as well as at 6 GeV for comparison.    The results are presented in Fig. \ref{Fig5}.

At 6 GeV, the fraction of particles crossing the selected energy-loss
threshold remains small, whereas at FCC-ee energies it reaches values of
approximately 10 \%. These particles are not necessarily lost from the machine; determining the actual loss rate would require a full lattice
simulation including the machine-specific momentum acceptance and
multi-turn beam dynamics.

It should be noted that the dependence in Fig. \ref{Fig5} is influenced not only by the radiation physics but also by the fraction of particles that undergo channeling. This fraction depends on the particle charge sign (compare Fig. \ref{Fig15}) as well as on the critical channeling angle and the ratio of the bending radius to its critical value, both scaling with the energy as described by Eqs. (\ref{Eq01}-\ref{Eq02}). For instance, the gradual decrease in radiation losses for electrons above $\sim$ 100 GeV is caused by the bending radius approaching its critical value, which reduces the channeling fraction.

These simulation results demonstrate that even a single passage through a bent crystal at high energies can substantially modify a significant amount of charged particle trajectories in an accelerator. Consequently, accurate simulation of radiation losses is essential for the design of crystal-based systems at future colliders.

In contrast, radiative energy losses in the crystal are far less important for muons, whereas deflection efficiency, decay-induced backgrounds and integration into a collider protection system become central issues. It nevertheless illustrates that our modular accelerator–crystal simulation framework can be extended to other lepton machines.

\section{Conclusion}
\label{conclusion}

A simulation model for crystal-based beam extraction based on the \textbf{BDSim} simulation toolkit and the Geant4 \textbf{G4ChannelingFastSimModel} and \textbf{G4BaierKatkov} models to simulate channeling and radiation processes in oriented crystals, respectively, has been developed. A new bent crystal accelerator component compatible with these models was created and integrated into a complete DESY II Booster Synchrotron accelerator model, including a full map of accelerator components along the ring, essential beam dynamics such as betatron and synchrotron oscillations, and radiation losses in the crystal.

The simulation successfully demonstrated the efficient extraction -- up to $\sim$ 28 \% -- of a monoenergetic 6 GeV electron beam with excellent emittance, a high charge of 23 pC, and a low energy spread of 0.008. The extracted beam fraction accounts for $\sim$ 1.4 \% of the total beam population, leaving $\sim$ 95 \% of the beam unaffected. Therefore, this parasitic extraction method offers a high-quality beam for diverse fixed-target applications without disturbing primary operations of the accelerator.

This simulation approach provides a comprehensive framework to design the applications of crystal-based extraction and collimation techniques in both current and future electron, positron and muon synchrotrons, synchrotron light sources, linacs and lepton colliders. \textbf{G4ChannelingFastSimModel} makes it possible to incorporate various coherent effects of charge particle deflection in an oriented crystal. \textbf{G4BaierKatkov} allows one to simulate the radiation energy loss being especially important at future collider energies.

The potential applications of our model include crystal-based extraction to supply fixed-target experiments with high-intensity, monoenergetic multi-GeV electron beams at existing accelerators and light sources worldwide. Furthermore, it could facilitate crystal-based electron/positron collimation at future lepton colliders, as well as extraction enabling  ultra-high energy fixed-target experiments beyond the collider physics. In addition, crystals can be exploited as intense sources of gamma radiation and secondary particles, such as positrons.

\section*{Acknowledgements}

A. Sytov was supported by the European Commission (TRILLION, GA. 101032975) during the original work presented in this paper. We acknowledge partial support of the INFN through the Geant4INFN project.

\section*{Code and Data availability}

The implementation of \textbf{CrystalDeflector} and \textbf{CrystalDeflectorConstructor}, together with the BDSim user-defined Physics List, are available for download in the
GitHub repository \citep{GITHUB}. The setup of crystal parameters in the BDSim macro is described below in Appendix \ref{appendix}. The DESY II Booster lattice is not publicly available due to DESY internal restrictions; however, users can build their own accelerator applications using the released code.

\appendix
\section{Setting up of the CrystalDeflector component in the BDSim macro}
\label{appendix}

The following macro line configures the \textbf{CrystalDeflector} component:\\
\texttt{CR1:usercomponent, userTypeName="crystaldeflector", l=0.4125, xsize=1*cm, ysize=1*cm, materialThickness=0.175*mm, offsetX=-1.126046*cm, offsetY=0*mm, axisX=0.000, axisY = -0.00097, axisZ = 0., horizontalWidth=1*m, material="G4\_Si", vacuumMaterial="vacuum", userParameters="crystalRegion:crystal1, crystalBendingAngle:0.00175, crystalLattice:(111), colour:decapole, radiationModel:true";}
\\ \\
In this setup:

\begin{itemize}
 \item \textbf{CR1} is the component's name.

 \item \textbf{userTypeName} identifies the component type as "crystaldeflector".

 \item \textbf{l}, \textbf{xsize}, \textbf{ysize}, and \textbf{materialThickness} define the dimensions and thickness of the crystal in the G4Box.

 \item \textbf{offsetX}, \textbf{offsetY}, and \textbf{axisX}, \textbf{axisY}, \textbf{axisZ} define the component center position and orientation in the beamline.

 \item \textbf{horizontalWidth} defines the vacuum space of the component around the crystal.

 \item \textbf{material} specifies the crystal material (e.g., silicon), and \textbf{vacuumMaterial} sets the surrounding vacuum.

 \item \textbf{userParameters} contain specific parameters for the crystal, such as the G4Region name for crystal channeling, bending angle, crystallographic orientation, visualization color, and radiation model activation.

\end{itemize}

For more information please refer the user's guide \citep{GITHUB}.


\begin{thebibliography}{00}


\bibitem{Tsyganov} E.N. Tsyganov, Some aspects of the mechanism of a charge particle penetration through a monocrystal. Fermilab TM-682 (1976). \url{https://lss.fnal.gov/archive/test-tm/0000/fermilab-tm-0682.shtml}

\bibitem[Afonin(2012)]{U70} A.G. Afonin et al., Observation and comparative analysis of proton beam extraction or collimation by different planar channels of a bent crystal. Phys. Rev. ST Accel. Beams \textbf{15}, 081001 (2012).
\url{https://doi.org/10.1103/PhysRevSTAB.15.081001}

\bibitem[Fliller(2005)]{RHIC} R.P. Fliller et al.
RHIC crystal collimation.
Nucl. Instr. and Meth. in Phys. Res. B \textbf{234} (1-2), 47 (2005).
\url{https://doi.org/10.1016/j.nimb.2005.03.004}

\bibitem[Scandale(2012)]{UA92} W. Scandale et al., Strong reduction of the off-momentum halo in crystal assisted collimation of the SPS beam. Phys. Lett. B \textbf{714}, 231 (2012)
\url{https://doi.org/10.1016/j.physletb.2012.07.006}

\bibitem[Carrigan(2002)]{Tevatron} R.A. Carrigan Jr. et al., Beam extraction studies at 900 GeV using a channeling crystal. Phys. Rev. ST Accel. Beams \textbf{5}, 043501 (2002).
\url{https://doi.org/10.1103/PhysRevSTAB.5.043501}

\bibitem[Scandale(2016)]{LHC} W. Scandale et al. Observation of channeling for 6500 GeV/c protons in the crystal assisted collimation setup for LHC. Phys. Lett. B \textbf{758}, 129 (2016). \url{https://doi.org/10.1016/j.physletb.2016.05.004}

\bibitem[Redaelli(2021)]{LHCion} S. Redaelli et al., 
First observation of ion beam channeling in bent crystals at multi-TeV energies. Eur. Phys. J. C \textbf{81}, 142 (2021).
\url{https://doi.org/10.1140/epjc/s10052-021-08927-x}

\bibitem[Scandale(2013)]{SPS2} W. Scandale et al., Optimization of the crystal assisted collimation of the SPS beam. Phys. Lett. B \textbf{726}, 182-186 (2013). \url{https://doi.org/10.1016/j.physletb.2013.08.028}

\bibitem[Sytov(2022)]{DESYII} A. Sytov et al.  First design of a crystal-based extraction of 6 GeV electrons for
the DESY II Booster Synchrotron. Eur. Phys. J. C
 \textbf{82}, 197 (2022).\url{https://doi.org/10.1140/epjc/s10052-022-10115-4}

\bibitem[Diener(2019)]{Diener} R. Diener et al. The DESY II test beam facility. Nucl. Instrum. Methods Phys. Res. A \textbf{922}, 265 (2019). \url{https://doi.org/10.1016/j.nima.2018.11.133}

\bibitem[Nevay(2020)]{BDSim1} L.J. Nevay et al. BDSIM: An accelerator tracking code with particle–matter interactions. Comp. Phys. Comm. \textbf{252}, 107200 (2020). \url{https://doi.org/10.1016/j.cpc.2020.107200}

\bibitem[Nevay(2018)]{BDSim2} L. Nevay, A. Abramov, S. T. Boogert, L.C. Deacon, H. Garcia-Morales, S. M. Gibson, R.Kwee-Hinzmann, W. Shields, J. Snuverink, and S. Walker. BDSIM: Automatic Geant4 models of accelerators, in: Proc. ICFA Mini-Workshop on Tracking for Collimation, vol. 45, CERN, Geneva, Switzerland, (2018) \url{http://dx.doi.org/10.23732/CYRCP-2018-002.45}

\bibitem[Nevay(2024)]{BDSim3} BDSim User's Guide \url{https://bdsim-collaboration.github.io/bdsim/sphinx/index.html}

\bibitem[Agostinelli(2003)]{Geant41} S. Agostinelli et al., GEANT4-A Simulation Toolkit. Nucl. Instrum. Meth. A \textbf{506}, 250-303 (2003). \url{http://dx.doi.org/10.1016/S0168-9002(03)01368-8}

\bibitem[Allison(2016)]{Geant42} J. Allison et al., Recent developments in Geant4. Nucl. Instrum. Meth. A \textbf{835}, 186-225 (2016). \url{https://doi.org/10.1016/j.nima.2016.06.125}

\bibitem[Enrico(2014)]{G4Channeling}  E. Bagli, M. Asai, D. Brandt, A. Dotti, V. Guidi, D. H. Wright. A model for the interaction of high-energy particles in straight and
bent crystals implemented in Geant4. Eur. Phys. J. C 74, 2996
(2014). \url{https://doi.org/10.1140/epjc/s10052-014-2996-y}

\bibitem[Broggi(2025)]{BDSimCollimation} G. Broggi, A. Abramov, M. Boscolo, R. Bruce, D. Mirarchi, S. Redaelli. First studies of crystal collimation for the FCC-ee. Nucl. Instrum. Methods Phys. Res. A \textbf{1076}, 170479 (2025). \url{https://doi.org/10.1016/j.nima.2025.170479}

\bibitem[Mazzolari(2020)]{CSS} A. Mazzolari, A. Sytov, L. Bandiera, G. Germogli, M. Romagnoni, E. Bagli, V. Guidi, V. V. Tikhomirov, D. De Salvador, S. Carturan, C. Durigello, G. Maggioni, M. Campostrini, A. Berra, V. Mascagna, M. Prest, E. Vallazza, W. Lauth, P. Klag, M. Tamisari. Broad angular anisotropy of multiple scattering in a Si crystal. Eur. Phys. J. C \textbf{80}, 63 (2020). \url{https://doi.org/10.1140/epjc/s10052-019-7586-6}

\bibitem[Kumakhov(1976)]{Kumakhov} M.A. Kumakhov, On the theory of electromagnetic radiation of charged particles in a crystal Phys. Lett. A 57(1), 17 (1976).
\url{https://doi.org/10.1016/0375-9601(76)90438-2}

\bibitem[Baryshevsky(1980)]{Baryshevsky} V.G. Baryshevsky, A.O. Grubich, I.Ya. Dubovskaya. Generation of $\gamma$-quanta by channeled particles in the presence of a variable external field. Phys. Lett. A \textbf{77}, 61-64 (1980). \url{https://doi.org/10.1016/0375-9601(80)90637-4}

\bibitem[Baier(1998)]{Baier} V. N. Baier, V. M. Katkov, and V. M. Strakhovenko, Electro-magnetic Processes at High Energies in Oriented Single Crystals (World Scientific, Singapore, 1998).

\bibitem[Ferretti(1950)]{Ferretti} B. Ferretti, Sulla "Bremsstrahlung" nei cristalli. Nuovo Cimento \textbf{7}, 118 (1950). \url{https://doi.org/10.1007/BF02781144}

\bibitem[Uberall(1956)]{Uberall} H. Überall, High-energy interference effect of bremsstrahlung and pair production in crystals, Phys. Rev. \textbf{103}, 1055 (1956). \url{https://doi.org/10.1103/PhysRev.103.1055}

\bibitem[(1972)]{Ter-Mikaelian} M.L. Ter-Mikaelian, High-Energy Electromagnetic Processes in
Condensed Media (Wiley, New York, 1972).

\bibitem[Mirarchi(2018)]{Sixtrack} D. Mirarchi, S. Redaelli, W. Scandale, Crystal implementation in SixTrack for proton beams, in: Proc. of the ICFA Mini-Workshop on Tracking for Collimation, 91-108 (2018). \url{http://dx.doi.org/10.23732/CYRCP-2018-002.91}

\bibitem[Skordis(2018)]{SixtrackFluka} E. Skordis, V. Vlachoudis, R. Bruce, F. Cerutti, A. Ferrari, A. Lechner, A. Mereghetti, P.G. Ortega, S. Redaelli, D. Sinuela Pastor. FLUKA coupling to sixtrack, in: Proc. of the ICFA Mini-Workshop on Tracking for Collimation,  17–25 (2018). \url{http://dx.doi.org/10.23732/CYRCP-2018-002.17}

\bibitem[Sytov(2023)]{Sytov} A. Sytov et al. Geant4 simulation model of electromagnetic processes in oriented crystals for accelerator physics. Journal of the Korean Physical Society \textbf{83}, 132–139 (2023). \url{https://doi.org/10.1007/s40042-023-00834-6}

\bibitem[Sytov(2019)]{CRYSTALRAD} A. I. Sytov, V. V. Tikhomirov, and L. Bandiera. Simulation code for modeling of coherent effects of radiation generation in oriented crystals. Phys. Rev. Acc. and Beams \textbf{22}, 064601 (2019).
\url{https://doi.org/10.1103/PhysRevAccelBeams.22.064601}

\bibitem[Sytov(2023)]{G4Ch} Channeling Fast Simulation Model User's Guide, Geant4 Physics Reference Manual, \url{https://geant4-userdoc.web.cern.ch/UsersGuides/PhysicsReferenceManual/html/solidstate/channeling/channeling_fastsim.html}

\bibitem[Sytov(2025)]{Riccardo} R. Negrello, L. Bandiera, N. Canale, P. Fedeli, V. Guidi, V.V. Haurylavets, A. Mazzolari, G. Paternò, M. Romagnoni, V.V. Tikhomirov, A. Sytov, A novel tool for advanced analysis of Geant4 simulations of charged particles interactions in oriented crystals. NIM A 1074, 170277 (2025). \url{https://doi.org/10.1016/j.nima.2025.170277}

\bibitem[Mazzolari(2025)]{PRL2025} A. Mazzolari, H. Backe, L. Bandiera, N. Canale, D. De Salvador, P. Drexler, V. Guidi, P. Klag, W. Lauth, L. Malagutti, R. Negrello, G. Paternò, M. Romagnoni, F. Sgarbossa, A. Sytov, V. Tikhomirov, and D. Valzani. Observation of Fine Structure in Channeling of Particles in Bent Crystals. Phys. Rev. Lett. 135, 205002 (2025). \url{https://doi.org/10.1103/t4ms-hmdd}

\bibitem[Canale(2025)]{Nicola} N. Canale, M. Romagnoni, A. Sytov, F. Alharthi, S. Bertelli, S. Carsi, I. Chaikovska, R. Chehab, D. De Salvador, P. Fedeli, V. Guidi, V. Haurylavets, G. Lezzani, L. Malagutti, S. Mangiacavalli, A. Mazzolari, P. Monti-Guarnieri, R. Negrello, G. Paternò, L. Perna, M. Prest, G. Saibene, A. Selmi, F. Sgarbossa, M. Soldani, V.V. Tikhomirov, D. Valzani, E. Vallazza, G. Zuccalà, L. Bandiera. Coherent radiation in axially oriented industrial-grade tungsten crystals: A viable path for innovative $\gamma$-ray and positron sources. Nucl. Instrum. Methods Phys. Res. A \textbf{1075}, 170342 (2025). \url{https://doi.org/10.1016/j.nima.2025.170342}

\bibitem[Lindhard(2025)]{Lindhard}  J. Lindhard, Influence of crystal lattice on motion of energetic charged particles, Mat. Fys. Medd. K. Dan. Vidensk. Selsk \textbf{34}, 64 (1965). \url{http://gymarkiv.sdu.dk/MFM/kdvs/mfm%2030-39/mfm-34-14.pdf}

\bibitem[Grinenko(1991)]{Axial1} A.A. Grinenko, N.F. Shul’ga, Turning a beam of high-energy charged particles by means of scattering by atomic rows of a curved crystal. J. Exp. Theor. Phys. Lett. \textbf{54}, 524 (1991). \url{http://jetpletters.ru/ps/1265/article_19143.shtml}

\bibitem[Shul’ga(1995)]{Axial2} N.F. Shul’ga, A.A. Greenenko, Multiple scattering of ultrahigh-
energy charged particles on atomic strings of a bent crystal. Phys. Lett. B \textbf{353}, 373 (1995). \url{https://doi.org/10.1016/0370-2693(95)00496-8}

\bibitem[Mazzolari(2014)]{rechanneling} A. Mazzolari et al., Steering of a Sub-GeV Electron beam through planar channeling enhanced by rechanneling. Phys. Rev. Lett. \textbf{112}, 135503 (2014). \url{https://doi.org/10.1103/PhysRevLett.112.135503}

\bibitem[Sytov(2017)]{rechanneling2} A. Sytov et al., Steering of sub-GeV electrons by ultrashort Si and Ge bent crystals. Eur. Phys. J. C \textbf{77}, 901 (2017).
\url{https://doi.org/10.1140/epjc/s10052-017-5456-7}

\bibitem[Tikhomirov(2018)]{MVROC} V. Tikhomirov, Multiple volume reflection from different planes inside one bent crystal. Phys. Lett. B \textbf{655}, 217 (2007). \url{https://doi.org/10.1016/j.physletb.2007.09.049}

\bibitem[Tikhomirov(2018)]{MVROC2} V. Guidi, A. Mazzolari, and V. Tikhomirov, On the observation of multiple volume reflection from different planes
inside one bent crystal. J. of Appl. Phys. \textbf{107}, 114908 (2010). \url{https://doi.org/10.1063/1.3407526}

\bibitem[Bandiera(2018)]{PRL2018}  L. Bandiera, V. V. Tikhomirov, M. Romagnoni, N. Argiolas, E. Bagli, G. Ballerini, A. Berra, C. Brizzolari, R. Camattari, D. De Salvador, V. Haurylavets, V. Mascagna, A. Mazzolari, M. Prest, M. Soldani, A. Sytov, and E. Vallazza. Strong Reduction of the Effective Radiation Length in an Axially Oriented Scintillator Crystal. Phys. Rev. Lett. \textbf{121}, 021603 (2018). \url{https://doi.org/10.1103/PhysRevLett.121.021603}

\bibitem[Bandiera(2012)]{PRA2012} V. Guidi, L. Bandiera, V. Tikhomirov. Radiation generated by single and multiple volume reflection of ultrarelativistic electrons and positrons in bent crystals. Phys. Rev. A \textbf{86}, 042903 (2012). \url{https://doi.org/10.1103/PhysRevA.86.042903}

\bibitem[Scandale(2009)]{MVROCCERN} W. Scandale et al., First observation of multiple volume reflection by different planes in one bent silicon crystal for high-energy protons. Phys. Lett. B \textbf{682}, 274–277 (2009), \url{https://doi.org/10.1016/j.physletb.2009.11.026}

\bibitem[Sytov(2022)]{GITHUB} G4ChannelingFastSimModel BDSim User's Guide \url{https://github.com/asytov00/G4ChannelingFastSimModelBDSim}

\bibitem[Handbook(2023)]{Applications1} Chao, A. W., Mess, K. H., Tigner, M., Zimmermann, F. (Eds.). (2023). Handbook of Accelerator Physics and Engineering (3rd ed.). World Scientific.

\bibitem[Dafna(2022)]{Dafna} M. Garattini, D. Annucci, O. R. Blanco-García, P. Gianotti, S. Guiducci, A. Liedl, M. Raggi and P. Valente, Crystal slow extraction of positrons from the Frascati DA$\Phi$NE collider, Phys. Rev. Accel. Beams \textbf{25}, 033501 (2022). \url{https://doi.org/10.1103/PhysRevAccelBeams.25.033501}

\bibitem[ApplicationsDESY(2019)]{ApplicationsDESY} R. Diener, J. Dreyling-Eschweiler, H. Ehrlichmann, I.M. Gregor, U. Kötz, U. Krämer, N. Meyners, N. Potylitsina-Kube, A. Schütz, P. Schütze, M. Stanitzki, The DESY II test beam facility. Nucl. Instrum. Methods Phys. Res. A \textbf{922}, 265 (2019). \url{https://doi.org/10.1016/j.nima.2018.11.133}

\bibitem[Kyryllin(2025)]{Muon} I.V. Kyryllin, M. Romagnoni, N. Canale, P. Fedeli, V. Guidi, L. Malagutti, A. Mazzolari, R. Negrello, G. Paternò, A. Sytov, L. Bandiera. Crystal assisted steering of muons and antimuons at the muon collider. Nucl. Instrum. Methods Phys. Res. A \textbf{1075}, 170385 (2025). \url{https://doi.org/10.1016/j.nima.2025.170385}

\bibitem[Agapov(2026)]{beyondFCC} I. Agapov, E. E. Alp, K. Andre, S. Antipov, A. Apyan, G. Arduini, L. Bandiera, W. Bartmann, H. Bartosik, M. Benedikt, S. Bettoni, J. M. Byrd, M. Calviani, A. Camper, C. Carli, S. Casalbuoni, A. Chance, P. Craievich, P. Crivelli, B. Dalena, M. Dickmann, M. Doser, I. Drebot, C. Duchemin, K. Dupraz, A. Frasca, S. J. Freeman, F. Gunsing, J. Jäckel, B. King, M. W. Krasny, A. Lechner, C. C. Lindstrøm, A. Mazzolari, C. Milardi, E. Musa, R. Negrello, F. Nguyen, K. Oide, Y. Papaphilippou, G. Paternò, V. Petrillo, K. Piotrzkowski, B. Rienäcker, G. Schnell, C. Schroer, I. Schulthess, L. Serafini, V. Shiltsev, M. Stampanoni, A. Variola, T. Watson, H.-U. Wienands, M. Wing, F. Zimmermann, Other science opportunities at the FCC-ee, Eur. Phys. J. Plus \textbf{141}, 271 (2026). \url{https://doi.org/10.1140/epjp/s13360-026-07399-w}

\bibitem[Baryshevsky(1990)]{BaryshevskySpin} V.G. Baryshevsky, Spin rotation and depolarization of relativistic particles traveling through a crystal. Nucl. Instrum. Methods Phys. Res. Sect. B \textbf{44}(3), 266–272 (1990). \url{https://doi.org/10.1016/0168-583x(90)90638-b}

\bibitem[Tikhomirov(1994)]{TikhomirovSpin} V.V. Tikhomirov, To the possibility to observe positron magnetic moment variation under the propagation through crystals. Sov. Yad. Phys. \textbf{57}, 2302 (1994).

\bibitem[Tikhomirov(1993)]{TikhomirovPolar} V.V. Tikhomirov, Possibility of observing radiative self-polarization and the production of polarized e+e- pairs in crystals at accessible energies. JETP Lett \textbf{58}, 168-171 (1993).

\bibitem[Lohmann(1994)]{CBMainz} D. Lohmann, J. Peise, J. Ahrens, I. Anthony, H.-J. Arends, R. Beck, R. Crawford, A. Hünger, K.H. Kaiser, J.D. Kellie, Ch. Klümper, P. Krahn, A. Kraus, U. Ludwig, M. Schumacher, O. Selke, M. Schmitz, M. Schneider, F. Wissmann, S. Wolf. Linearly polarized photons at MAMI (Mainz). Nucl. Instr. and Meth. in Phys. Res. A \textbf{343}, 494-507 (1994). \url{https://doi.org/10.1016/0168-9002(94)90230-5}

\bibitem[Adhikari(2021)]{CBJLAB} S. Adhikari et al., The GlueX beamline and detector. Nucl. Instrum. Methods Phys. Res. A \textbf{987}, 164807 (2021). \url{https://doi.org/10.1016/j.nima.2020.164807}


\bibitem[ApplicationsMainz(2015)]{ApplicationsMainz} D. Lietti, H. Backe, E. Bagli, L. Bandiera, A. Berra, S. Carturan, D. De Salvador, G. Germogli, V. Guidi, W.
Lauth, A. Mazzolari, M. Prest, E. Vallazza, The experimental setup of the Interaction in Crystals for Emission of RADiation collaboration at Mainzer Mikrotron: Design, commissioning, and tests.  Rev. of Sci. Instr. \textbf{86}, 045102 (2015). \url{http://dx.doi.org/10.1063/1.4916367}

\bibitem[Bandiera(2015)]{PRL2015} L. Bandiera, E. Bagli, G. Germogli, V. Guidi, A. Mazzolari, H. Backe, W. Lauth, A. Berra, D. Lietti, M. Prest, D. De Salvador, E. Vallazza, V. Tikhomirov. Investigation of the Electromagnetic Radiation Emitted by Sub-GeV Electrons in a Bent Crystal. Phys. Rev. Lett. \textbf{115}, 025504 (2015). \url{https://doi.org/10.1103/PhysRevLett.115.025504}

\bibitem[Sytov(2021)]{EPJC2021} L. Bandiera, A. Sytov, D. De Salvador, A. Mazzolari, E. Bagli, R. Camattari, S. Carturan, C. Durighello, G. Germogli, V. Guidi, P. Klag, W. Lauth, G. Maggioni, V. Mascagna, M. Prest, M. Romagnoni, M. Soldani, V. V. Tikhomirov, E. Vallazza. Investigation on radiation generated by sub-GeV electrons in ultrashort silicon and germanium bent crystals. Eur. Phys. J. C \textbf{81}, 284 (2021). \url{https://doi.org/10.1140/epjc/s10052-021-09071-2}

\bibitem[Wienands(2017)]{SLAC2017} U. Wienands, S. Gessner, M.J. Hogan, T.W. Markiewicz, T. Smith, J. Sheppard, U.I. Uggerhøj, J.L. Hansen, T.N. Wistisen, E. Bagli, L. Bandiera, G. Germogli, A. Mazzolari, V. Guidi, A. Sytov, R.L. Holtzapple, K. McArdle, S. Tucker, B. Benson. Channeling and radiation experiments at SLAC. Nucl. Instr. and Meth. in Phys. Res. B. \textbf{402}, 11-15 (2017). \url{http://dx.doi.org/10.1016/j.nimb.2017.03.097}

\bibitem[Wienands(2019)]{SLAC2019} U. Wienands, S. Gessner, M. J. Hogan, T. Markiewicz, T. Smith, J. Sheppard, U. I. Uggerhøj, C. F. Nielsen, T. Wistisen, E. Bagli, L. Bandiera, G. Germogli, A. Mazzolari, V. Guidi, A. Sytov, R. L. Holtzapple, K. McArdle, S. Tucker, B. Benson, and The SLAC T513–E212–T523 Collaboration. Channeling and radiation experiments at SLAC. Int. J. Mod. Phys. \textbf{34}, 1943006 (2019). \url{https://dx.doi.org/10.1142/S0217751X19430061}

\bibitem[Kaplin(2019)]{CU2} V. V. Kaplin, S. V. Plotnikov, and S. A. Vorobiev, Radiation by charged particles in deformed crystals, phys. stat. sol. (b) 97, K89 (1980). \url{https://doi.org/10.1002/pssb.2220970237}

\bibitem[Tabrizi(2007)]{CU3} M. Tabrizi, A. V. Korol, A. V. Solovyov, and W. Greiner, Feasibility of an Electron-Based Crystalline Undulator, Phys. Rev. Lett. 98, 164801 (2007). \url{https://doi.org/10.1103/PhysRevLett.98.164801}

\bibitem[Korol(2014)]{CU4} A.V. Korol, A.V. Solov’yov, W. Greiner, Channeling and Radiation in Periodically Bent Crystals (Springer Heidelberg, New York, Dordrecht, London, 2014).

\bibitem[PositronSource(2022)]{PositronSource} L. Bandiera, L. Bomben, R. Camattari, G. Cavoto, I. Chaikovska, R. Chehab, D. De Salvador, V. Guidi, V. Haurylavets, E. Lutsenko, V. Mascagna, A. Mazzolari, M. Prest , M. Romagnoni, F. Ronchetti, F. Sgarbossa, M. Soldani, A. Sytov, M. Tamisari, V. Tikhomirov, E. Vallazza. Crystal-based pair production for a lepton collider positron source. Eur. Phys. J. C, \textbf{82}, 699 (2022). \url{https://doi.org/10.1140/epjc/s10052-022-10666-6}

\bibitem[Alharthi(2025)]{PositronSource2} F. Alharthi, I. Chaikovska, R. Chehab, V. Mytrochenko, Y. Wang, Y. Zhao, L. Bandiera, N. Canale, V. Guidi, L. Malagutti, A. Mazzolari, R. Negrello, G. Paternò, M. Romagnoni, A. Sytov, D. Boccanfuso, A.O.M. Iorio, S. Bertelli, M. Soldani. FCC-ee positron source from conventional to crystal-based. Nucl. Instrum. Methods Phys. Res. A \textbf{1075}, 170412 (2025). \url{https://doi.org/10.1016/j.nima.2025.170412}

\end{thebibliography}

\end{document}